\documentstyle[12pt]{ioplppt}
\begin{document}
\baselineskip=18pt
\title{Analysis of ``Gauge Modes'' in Linearized Relativity}
\author{Richard A. Matzner, Mijan Huq, Alonso Botero, Dae Il Choi, 
Ullar Kask, Juan Lara, Steven Liebling, David Neilsen, Premana Premadi 
and Deirdre Shoemaker.}
\address{\em Department of Physics and Center for Relativity, \\
The University of Texas at Austin. Tx 78712-1081}
\begin{abstract}
By writing the complete set of $3 + 1$ (ADM) equations for linearized waves,
we are able to demonstrate the properties of the initial data and of the 
evolution of a wave problem set by Alcubierre and Schutz. We show that the
gauge modes and constraint error modes arise in a straightforward way
in the analysis, and are of a form which will be controlled in any well
specified convergent computational discretization of the differential
equations.
\end{abstract}
\maketitle
\section{Introduction}

The $3 + 1$ (ADM \cite{ADM}) form of the Einstein equations for vacuum may be
written:
\begin{eqnarray}
R + K^{2} - K_{cd} K^{cd} = 0 \\ \label{1.1} 
\nabla_a K^{a}_{~b} - \nabla_b K = 0\\ \label{1.2}
\hat{\partial}_{o} g_{ab} = -2 \alpha K_{ab}\\ \label{1.3}
\hat{\partial}_{o} K_{ab} = -\nabla_{a} \nabla_{b} \alpha
+ \alpha [R_{ab} - 2K_{al}K^{l}_{~b} + KK_{ab}], \label{1.4}
\end{eqnarray}
where the derivative [2] $\hat{\partial}_{o}$ is defined:
\begin{eqnarray}
\partial_{o} = \hat{\partial}_{o} + \pounds_{\beta}.
\end{eqnarray}

Here $\alpha$ is the lapse function, which relates the coordinate time between
time instants to the proper time interval, $\beta^i$ is a shift vector
describing changes in the coordinization at one time and 
$\pounds_{\beta}$ is the Lie derivative along $\beta^i$. Because they are 
kinematical choices, we can, for some (perhaps small) interval, set $\alpha$
and $\beta^{i}$ arbitrarily. Here we take $\alpha = 1$, $\beta^{i} = 0$,
following Alcubierre and Schutz, \cite{alcubierre.etal} who posed the problem 
we now define. We use the overdot and $\partial_{o}$ interchangeably
below to indicate the time derivative.

In Eqs.~(1 - 4), all of the geometrical variables are 3-tensors.
$g_{ab}$ is the spatial 3 metric; $K_{ab}$ is a spatial tensor
(with $\alpha =1$, $\beta^{i} = 0,$ equal to
$-\frac{1}{2} \dot{g}_{ab})$; $K$ is the 3-trace of $K_{ab}$; $R_{ab}$ is the 3-Ricci
tensor obtained from $g_{ab}$, and $R$ is the trace of $R_{ab}$.
Linearizing the equation around flat space $(g_{ab} = \delta_{ab} + 
h_{ab};
K_{ab} \; \; \mbox{``small''})$ and introducing the traceless variables
$\stackrel{\circ}{K}_{ab} = K_{ab} - \frac{1}{3} \delta_{ab}K,$ and 
$\stackrel{\circ}{h}_{ab} = h_{ab} - \frac{1}{3} h_{cd}\delta^{cd} \delta_{ab},$ we have
\begin{eqnarray}
R = 0 \\
\stackrel{\circ}{K} {^{a}}_{b,a} = \frac{2}{3} K,_{b} \\
\partial_{o}\stackrel{\circ}{h}_{ab} = -2\stackrel{\circ}{K}_{ab} \\
\dot{h} = -2 {K} \\
\partial_{o}\stackrel{\circ}{K}_{ab} = R_{ab} - \frac{1}{3} \delta_{ab} R \\
\dot{K} = R.
\end{eqnarray}
Here $R_{ab}$, the 3-Ricci tensor, is (for the linearized case):
\begin{eqnarray}
R_{ab}=\frac{1}{2}
\delta^{mn} 
(-h_{ab, mn} + h_{an, bm} + h_{bn, am} -h_{mn, ab})
\end{eqnarray}
where the comma now denotes the partial derivative.

Formally, Einstein equations are a constrained hyperbolic system. Solution
of the constraint equations (1 - 2) [or their linearized version 
(6 - 7)] will be preserved analytically in the evolution of the dynamical
equations (3 - 4) [or their linearization]. 
By setting $\alpha = 1, \beta^{i}= 0$,  
we choose not to consider the admissible 
possibility of perturbations in $\alpha$ and $\beta^{i}$ of the same order as 
$h$. This is a coordinate choice (gauge choice).
Further, there is the possibility to specify the gauge more completely at the 
perturbation level in a way that does not change $\alpha, \beta^i$. 
Alcubierre and Schutz [3], for instance, choose a transverse traceless gauge, 
$\delta^{mn}h_{mn} = 0$ and $\delta^{mn}\partial_{m}h_{ni} = 0.$ Such a choice 
can be accomplished by a coordinate transformation at the original time; 
analytically this choice is preserved with $\alpha = 1, \beta^{i} = 0$
in the evolution. Gauges like this represent additional
restrictions on the behavior of the solution; however they are not necessary
to have a physically meaningful evolution. We shall see below that
even in this linearized case, setting such gauges in computation is
delicate.

\bigskip

\section{``Gauge Modes''}

Alcubierre and Schutz wrote the linearized evolution equations (8 - 11)
in second order form without separating out the trace:
\begin{eqnarray}
(-\partial^{2}_{0} + \nabla^{2})h_{ij} = \delta^{mn}(\partial_{n}
\partial_{i}h_{jm} + \partial_{n}\partial_{j}h_{im} - 
\partial_{i}\partial_{j}h_{mm}).
\end{eqnarray}

With this form, Alcubierre and Schutz\cite{alcubierre.etal} discovered 
secular, non-propagating, modes in their numerical implementation. 
In their attempt to evolve transverse traceless waves they find offending
extra modes that violate the TT-gauge condition. Some of these modes appear
to violate the constraints.  By investigation of the  properties of
Eq.~(13), Alcubierre and Schutz
were able to show analytically the existence of modes
similar to those seen numerically.
\bigskip

\section{Preliminaries}

From Eq.~(12) for the linearized Ricci tensor, it is easy to obtain the
linearized 3-scalar $R$:
\begin{eqnarray}
R = - \nabla^{2}h +  h^{mn},_{mn}
\end{eqnarray}
and the linearized Hamiltonian constraint (6) sets this to zero.

A basic decomposition which we employ is:
\begin{eqnarray}
h_{mn} = h^{TT}_{mn} + h^{L}_{mn} + \frac{1}{3}\delta_{mn}h^{*}.
\end{eqnarray}

Here $h^{TT}_{mn}$ is transverse:
\begin{eqnarray}
h^{TTmn}{}_{,n} = 0,
\end{eqnarray}

and traceless:
\begin{eqnarray}
h^{TT}{^{n}}_{n} = 0.
\end{eqnarray}

The quantity $h^{L}_{mn}$ is longitudinal, and can be written:
\begin{eqnarray}
h^{L}_{mn} = v_{m,n} + v_{n,m}
\end{eqnarray}

for some vector $v_{m}$. Notice that $h^{L}_{mn}$ is not tracefree:
\begin{eqnarray}
h^{L}{_{m}}^{m} = 2 v_{m}{^{,m}}
\end{eqnarray}
(The construction (18 - 19) for $h^{L}{_{mn}}$ is similar to that given
by York\cite{york} except that we do not subtract the trace from 
$h^{L}{_{mn}}$). Additionally we posit an independent contribution to the 
trace, $h^{*}$.
At this stage we distinguish $\stackrel{\circ}{h}_{mn}$, which need not
be transverse, from $h^{TT}{_{mn}}$, which is transverse.

Firstly, notice that $h^{L}_{mn}$ annihilates the scalar curvature. This
is not surprising because such longitudinal components arise from classic
``gauge terms'' as in Eq.~(18). 

Secondly, notice that if we set (only) 
nonvanishing $h^{TT}_{mn}$ data, $R = 0$ becomes
\begin{eqnarray}
\nabla^{2}h^{*} = 0.
\end{eqnarray}
In this study we are concerned with computational relativity, for
``physically realistic'' disturbances. In particular, we will expect
fall-off ``at infinity'' (this can be accomplished by a mixed
outer boundary condition on the computational domain), 
and we exclude singularities in the interior
of the domain. The solution to Eq.~(20) then is $h = 0$. Hence, we expect
the trace, $h$, to arise only in the context of longitudinal components, 
$h^{L}_{mn}$, as in Eq.~(19).

\section{Linearized Data Setting and Evolution}

We continue the Alcubierre and Schutz analysis of the linearized case,
but take more careful note of the constraint equations in the analysis.
We will also give a complete data-setting analysis.

An immediate result from the Hamiltonian constraint $R = 0$ and from the 
second order form of the linearized evolution equations
(8 - 11) is:
\begin{eqnarray}
\partial_{o}^{2}h = 0 \; \; \; \; \; (!)
\end{eqnarray}
Hence the trace of the metric perturbation is solved by
\begin{eqnarray}
h = a (x^{i}) + b(x^{i})t,
\end{eqnarray}
if one enforces the Hamiltonian constraint to this order. Notice that since
$K = -\frac{1}{2}\dot{h}$, we have $K = -\frac{1}{2}b(x^{i})$. Notice
also that because $h$ can only arise from longitudinal modes, we have
found Alcubierre and Schutz's growing modes:
\begin{eqnarray}
h^{L}_{mn} = A^{L}_{mn} (x^{i}) + t B^{L}_{mn}(x^{i}),
\end{eqnarray}
However, also note that $B^{L}_{mn}$ is related to $K_{mn}$ and
is restricted by the momentum constraint.

Similarly writing the second-order equation for $\stackrel{\circ}{h}_{ab}$,
we find:
\begin{eqnarray}
(-\partial_{o}^{2} + \nabla^{2})\stackrel{\circ}{h}_{ij} & = & 2
\partial_{n}\partial_{({i}}\stackrel{\circ}{h}{_{j)}^{n}} - 
\frac{1}{3} \partial_{i} \partial_{j}h\\ \nonumber
& & -\frac{1}{3}(2 \partial_{n}\partial_{m}\stackrel{\circ}{h}{^{mn}} -
\frac{1}{3}\nabla^{2}h) \delta_{ij}.
\end{eqnarray}
The last term (in parenthesis, proportional to $\frac{1}{3}\delta_{ij}$)in 
Eq.~(24) is in fact part of the Ricci scalar, and again using the
Hamiltonian  constraint it equals $-\frac{1}{3}\nabla^{2}h \delta_{ij}$:
\begin{eqnarray}
(-\partial_{o}^{2} + \nabla^{2})\stackrel{\circ}{h}_{ij} =
2 \partial_{n}\partial_{(i}\stackrel{\circ}{h}{_{j)}^{n}} 
-\frac{1}{3} (\partial_{i}\partial_{j} + \delta_{ij} \nabla^{2})h.
\end{eqnarray}

Clearly, $TT$ data annihilates the righthand side of Eq.~(25), so 
$h^{TT}_{mn}$ satisfies the source free wave equation. It is also
straightforward to verify using Eq.~(18) that $h^{L}_{mn}$ annihilates
the sum of the spatial derivatives in Eq.~(25), so that
\begin{eqnarray}
\partial_{o}^{2} h^{L}_{mn} = 0,
\end{eqnarray}
consistent with the behavior we have already found for the trace. 
Thus this system is {\it simply hyperbolic} in the notation of reference
\cite{abrahams.etal}: disturbances travel at speed zero $(h^{L}{_{mn}})$, or
unity $(h^{TT}{_{mn}})$.
In order to proceed, we should also investigate the time derivative
of the right side of (24). Let us proceed by taking note of
one of the remarkable lessons learned from the hyperbolic analysis, for 
instance like that carried out in reference \cite{abrahams.etal}, that
one should consider higher temporal derivatives of the evolution
equations. For instance, an additional time derivative gives:
\begin{eqnarray}
(-\partial_{0}{^{2}} + \nabla^{2})\stackrel{\circ}{K}
_{ij} & = & \partial_{i}(\stackrel{\circ}{K}{_{j}^{n}}),_{n} +
\partial_{j} (\stackrel{\circ}{K}{^{n}_{i}}),_{n} \nonumber \\
& & -\frac{1}{3} \partial_{i}\partial_{j}K - \frac{1}{3}\delta_{ij}
\nabla^{2}K.
\end{eqnarray}

By the momentum constraint, (7) the first two terms on the right are equal;
each equal to $\frac{2}{3} K,_{ij}$.

Thus:
\begin{eqnarray}
(-\partial_{o}^{2} + \nabla^{2})\stackrel{\circ}{K}_{ij} & = &
\partial_{i}\partial_{j}K -\frac{1}{3}\nabla^{2}K \delta_{ij} \\
& = & \mbox {given function of spatial coordinates} \nonumber \\
& = & -\frac{1}{2} (\partial_{i} \partial_{j}b - \frac{1}{3}
\delta_{ij} \nabla^{2}b).  \nonumber 
\end{eqnarray}
Hence;
\begin{eqnarray}
\stackrel{\circ}{K}_{ij} & = & \stackrel{\circ}{K}_{ij} (\mbox{wave}) + \nabla^{-2} (\partial_{i}\partial_{j}K - \frac{1}{3}\nabla^{2}K\delta_{ij}).
\end{eqnarray}

Where $\nabla^{-2}$ is the Greens function with appropriate boundary conditions.
Furthermore, one can consider the temporal evolution of $\delta^{il}\stackrel
{\circ}{K}_{ij,l}$, the divergence of $\stackrel{\circ}{K}_{ij}$.

By reorganizing Eq.~(28) to have only the second time derivative on the left,
and taking the divergence, we obtain
\begin{eqnarray}
\partial_{o}^{2} \stackrel{\circ}{K}{^{ij}{_{,j}}} = 0,
\end{eqnarray}
which in principle allows
\begin{eqnarray}
\stackrel{\circ}{K}{^{ij}{_{,j}}} = c^{i}(x^{j}) + d^{i}(x^{j})t.
\end{eqnarray}

If we inspect Eq.~(28) again, however, we find that solutions consist of
solutions to the homogeneous wave equation (which, by Eq.~(30) must
be divergenceless) plus a ``particular'' solution which has {\it no}
time dependence, because the right hand side is time independent. Hence
$d_{i}(x^{j}) \equiv 0$ in Eq.~(30), and we may write
\begin{eqnarray}
\stackrel{\circ}{K}_{ij} = K^{TT}_{ij} + \nabla^{-2} 
(\partial_{i}\partial_{j}K
- \frac{1}{3} \delta_{ij}\nabla^{2} K). 
\end{eqnarray}

Consider an arbitrary longitudinal $K^{L}_{ij} = w_{i,j} + w_{j,i}.$ 
A general 3-tensor of this form does not satisfy the momentum constraint 
$K_{i}{^{j}}_{, j} = \delta_{i}^{j}K_{,j}$:
\begin{eqnarray}
w_{i},^{j}{_{,j}} + w^{j}_{,i,j} - 2w_{l}{^{,l}}{_{,i}} = 
w_{i}{^{,j}}{_{,i}} - w^{j}{_{,i,j}}
\end{eqnarray}
Hence in order to satisfy the momentum constraint we must have
\begin{eqnarray}
(w_{i,j} - w_{j,i})_{,k}\delta^{jk} = 0.
\end{eqnarray}
The simplest solution is to take $w_{l} = \psi_{,l},$ the gradient of a 
scalar. Although more general solutions may be possible, but likely excluded
by reasonable boundary conditions. With $w_{l} = \psi_{,l}$ we have 
\begin{eqnarray}
2\nabla^{2}\psi  = K. 
\end{eqnarray}
The corresponding $\stackrel{\circ}{K}{^{L}_{ij}}$ is 
\begin{eqnarray}
\stackrel{\circ}{K}{^{L}_{ij}} = 
2 \psi_{,ij} - \frac{2}{3} \delta_{ij}\nabla^{2}\psi = 
2 \psi_{,ij}
-\frac{1}{3} \delta_{ij}K.
\end{eqnarray}

\section{Transverse Data}
\label{transverse.data}

As an aside, we state how one sets $TT$ data. Obviously, if the data are set with a given (fixed) wave vector direction,
one can algebraically make the tensors $TT$. Also, if one wishes to 
Fourier decompose the data, then each Fourier component can be made transverse.
In general, one can use the following procedure.
The idea is to pose an arbitrary
traceless field $C_{mn}$, and compute its divergence.

Then, find the longitudinal component $L_{mn} = w_{m,n}+ w_{n,m}$ by
solving
\begin{eqnarray}
C_{m}{}^{n}{}_{,n} = \nabla_{n} 
(\nabla_{m}w^{n} + \nabla^{n}w_{m}), \\
\mbox {i.e.:} \nonumber \\
\nabla_{n} \nabla^{n}w_{m} 
+ \nabla_{m} (\nabla \cdot w) = C_{m}{^{n}}_{,n}.
\end{eqnarray}

The operator on the left is easily derived from the minimization of
$\nabla_{(m}w{_{n)}}\nabla^{(m}w{^{n)}}$ and so is strongly
elliptic, guaranteeing the existence of unique solutions. If, for instance one
postulates compact support for $C_{mn}$, one can solve this equation 
using a scalar potential field $\phi: w_{m} = \phi_{,m}$
\begin{eqnarray}
2 \nabla^{2} \phi_{,m} = C_{m}{^{n}}_{,n}, 
\end{eqnarray}
so that
\begin{eqnarray}
w_{m} = \phi_{,m} = \nabla^{-2} 
(\frac{1}{2} C_{m}{^{n}}_{,n}) \sim \frac{1}{r} + O (\frac{1}{r^{2}}).
\end{eqnarray}
The resulting longitudinal component $L_{mn}$ thus has behavior
$\sim O (r^{-2})$, and the transverse data now is: $C_{mn} - L_{mn},$
and no longer has compact support.

\section{Complete Data Setting}

Set $h^{TT}_{mn}$, $h^{L}_{mn} = v_{m,n}+ v_{n,m}$ with appropriate
locality (eg. compactness; at least ``fall-off at infinity). Then 
the trace $h$ is specified; cf. eqs. (19 - 20).

Set $K^{TT}_{mn}, K$ with appropriate locality. Then $K^{L}_{mn}$ is set by 
Eq.~(36) in terms of $\psi$ where $2 \nabla^{2} \psi = K$ and $K^{L}_{mn}$
satisfies the momentum constraint. The Hamiltonian constraint is maintained
by {\it any} $K^{L}_{mn}$, in particular one of form (36).

\section{Expected Numerical Behavior - Unconstrained Case}

From Eq.~(32) $\stackrel{\circ}{h}{_{ij}}$ satisfies
\begin{eqnarray}
\stackrel{\circ}{h}{^{H}_{ij}} & = & \stackrel{\circ}{h}{^{L}_{ij}}(t_{o}) -2 t \nabla^{-2}(\partial_{i}\partial_{j}K - \frac{1}{3}\delta_{ij}\nabla^{2}K)
 + {h}^{TT}_{ij}\mbox{(wave)}.
\end{eqnarray}
Note that $\stackrel{\circ}{h}{_{ij}}$ (wave) is transverse because its time
derivative
$-2 {K}^{TT}_{ij}$ is transverse.

Since we have specified the intermediate variable $\psi$ by $2 \nabla^{2} \psi = K$,
we may also write:

\begin{eqnarray}
\stackrel{\circ}{h}{_{ij}} = \stackrel{\circ}{h}{^{L}_{ij}} -2t (2 \partial_{i} \partial_{j} \psi
-\frac{1}{3} \delta_{ij}K) + h^{TT}_{ij}.
\end{eqnarray}
To proceed, let us suppose that we wish, as Alcubierre and Schutz did,
to set purely $TT$ data. 

In the Alcubierre-Schutz case, plane waves were set, so $TT$ data can be algebraically
enforced, and we are solving only
\begin{eqnarray}
\Box h^{TT}_{nm} = 0.
\end{eqnarray}
However, errors in setting $TT$ data, can lead to a nonzero longitudinal
part. Since we solve to zero the longitudinal component via an elliptic
equation in a second order scheme we expect the longitudinal error to be
second under small:
\begin{eqnarray}
|| \Delta h^{L}_{mn}|| \sim \left(\frac{\Delta x}{\lambda}\right)^{2} 
||h^{TT}_{mn}||
\end{eqnarray}
where $\Delta x$ is the discretization scale, and $\lambda$ 
is the typical scale of the $TT$ part. Then, since longitudinal terms grow as
$t$, one expects longitudinal contamination equal to the transverse
signal at
\begin{eqnarray}
\frac{t}{\lambda} \sim (\frac{\Delta x}{\lambda})^{-2}.
\end{eqnarray}

For typical discretizations, $(\frac{\Delta x}{\lambda})
\stackrel{\sim}{<} 10^{-2}$ at worst, which suggests evolutions for times
\begin{eqnarray}
t \sim 10^{4}\lambda
\end{eqnarray}
before the longitudinal signals are comparable to the transverse signal,
and another factor of order $O \left(||h^{TT}_{mn}||^{-1}\right)$ before the 
longitudinal mode violates the linearization criterion $\|h^L_{mn}\| \ll 1$. 
Unfortunately, a poorly controlled computational scheme could put error 
(noise) in the longitudinal mode at short scales (so that 
$\frac{\Delta x}{\lambda} \sim 1$).
In that case one would expect to see the longitudinal mode grow to the
amplitude of the transverse signal after only a few time steps;
in a poorly designed differencing approach these secular zone-to zone
oscillations would crash the program shortly thereafter. Notice that errors
in the trace $h$ will grow similarly to the $h^{L}{_{mn}}$ modes and with
potentially worse effect, because they violate the Hamiltonian constraint.
Without a close inspection of the Alcubierre-Schutz code, we cannot comment 
on why they found such rapid growth of non-$TT$ modes.

In a recent paper\cite{gundlach.etal}, Gundlach and Pullin point out a
mechanism for instabilities arising from the violation of the constraints
in a free evolution. They used perturbation analysis in double null 
coordinates on a Reissner-Nordstr\"{o}m background, and found that a free 
evolution led to exponentially growing gauge violating modes. Their results
can be taken to the flat space limit. In contrast to the Gundlach and Pullin 
result, our flat background analysis finds modes that grow at most linearly 
with time.  Choptuik\cite{choptuik} suggests that the presence of the $r=0$ 
singularity (persisting even as one takes the $M\rightarrow 0$ and 
$q\rightarrow 0$ limit) in \cite{gundlach.etal} results in exponentially 
growing modes rather 
than the linearly growing modes that we see in our analysis, which explicitly
assumed regularity.

\section{Constrained Evolution}
In view of Eq.~(20), violations of the Hamiltonian constraint lead to spurious 
$h^{*}$. This can be solved for $h^{*}$:
\begin{eqnarray}
h^{*} = \nabla^{-2} R.
\end{eqnarray}
This $h^{*}$ can then be subtracted from the solution at any particular
instant, reasserting the Hamiltonian constraint.

Gauge drift, which would arise from an error in setting exactly zero
longitudinal data, can be similarly suppressed. Section \ref{transverse.data}
shows how to remove longitudinal data from arbitrarily set data. This step can 
be carried out at any particular instant, reasserting the $TT$ requirement.

\section{Summary}

We have shown that gauge and constraint error modes arise in the analysis of
the 3+1 form of the Einstein equations for linearized waves. These modes are
shown to grow linearly in time and have a form that can be controlled in any
well specified convergent computational discretization of the evolution 
equations. By imposing the constraints on the free-evolution these modes may 
be supressed.
\section{Acknowledgments}

This work was supported by NSF grants PHY 9310082 and PHY 931852. Computer
time was supported by the High Performance Computing Facility,
The University of Texas at Austin.



\end{document}